# Scaling Invariance and Characteristics of the Fragments Cloud of Spherical Projectile Fragmentation upon High-Velocity Impact on a Thin Mesh Shield

N.N. Myagkov

*Institute of Applied Mechanics, Russian Academy of Sciences, Leningradsky Prospect 7, Moscow 125040, Russia*

**Abstract**

In the present paper we consider the problem of the fragmentation of an aluminum projectile on a thin steel mesh shield at high-velocity impact in a three-dimensional (3D) setting. The numerical simulations are carried out by smoothed particle hydrodynamics method applied to the equations of mechanics of deformable solids. Quantitative characteristics of the projectile fragmentation are obtained by studying statistics of the cloud of fragments. The considerable attention is given to scaling laws accompanying the fragmentation of the projectile. Scaling is carried out using the parameter *K* which defines the number of the mesh cells falling within the projectile diameter**.** It is found that the dependence of the critical velocity $V_c$ of fragmentation on the parameter *K* consists of two branches that correspond to two modes of the projectile fragmentation associated with the "small" and "large" aperture of the mesh cell. We obtain the dependences of the critical velocity $V_c$ on the projectile diameter and the mesh parameters for the both modes of the fragmentation. It is shown that the average cumulative mass distributions constructed at $V_c$ exhibit the property of scale invariance, splitting into two groups of distributions exactly corresponding to the modes of the projectile fragmentation. In each group, the average cumulative distributions show good coincidence in the entire mass region, moreover in the intermediate mass region the each group of distributions has a power-law distribution with an exponent τ different from that in the other group. The conclusion about the dependence of the exponent of the power-law distribution τ on the fragmentation mode is made.

## 1. Introduction

Dynamic fragmentation of solids caused by impact or explosion has been studied for years [1–25]. Typical experimental situations in which these phenomena take place correspond to collisions of heavy nuclei in atomic physics [3, 4, 19, 20], collision of macroscopic bodies [1, 12,

16] (including projectile impact on a massive target [1, 17, 18, 23, 25]), shell fragmentation upon explosion [13], and projectile fragmentation upon high-velocity perforation of a thin shield [2, 5, 6, 8-11, 21, 22].

A possible critical behavior during fragmentation was originally analyzed in the framework of a problem of nuclear collisions at moderate energies [3, 4, 19, 20] using an approach based on the similarity of the observed distribution of fragments and that predicted by well-known theories of critical phenomena such as liquid–vapor transition and percolation. Later, these methods were applied to studying fragmentation in mechanical systems [8-10, 12-18, 24]. The existence of a critical transition from damage to fragmentation was confirmed for mechanical systems of various types both experimentally [14, 15] and numerically [8-10, 11-13, 16–18].

The transition from damage to fragmentation (or the degree of fracture) can be characterized in different ways, including the average fragment mass, mass of maximum fragment, fluctuations of the mass of largest fragment, etc. [3, 4, 12, 13, 19, 20]. In systems with impact interaction, a control parameter is usually selected to be the impact velocity or energy [8-10, 12–13] or the total number of fragments [3, 4, 14].

Experiments [1,3, 13-16,19, 23,25] and numerical simulations [7, 10, 12, 13, 16, 24] showed that the mass distribution of the fragments can be represented by a power function

$$n(m) \sim m^{-\tau} \tag{1}$$

in some non-negligible range of fragments mass variation. It is known that the relation (1) is a necessary but not sufficient condition for the critical behavior in fragmentation. In case of percolation the exponent happens to be $\tau > 2.0$ in three dimensions [26]. Experiments on nuclear fragmentation yield values of $\tau$ consistent with this inequality [3,4], while experiments on fragmentation of brittle and plastic materials give values of $\tau$ both larger and smaller than 2.0 [7, 9, 10, 12-18, 23-25].

Experiments and numerical simulations have shown that there is a relationship between $\tau$ and effective dimension of the fragmented object, i.e. the exponent $\tau$ grows as the effective dimension increases [23]. Dependence of the exponent $\tau$ on the initial energy imparted to a fragmented object was obtained in experiments and numerical simulations (e.g., [11, 12, 15, 24]); it was found that the exponent $\tau$ increases with the initial energy. The authors of several papers (see, e.g., [24]) interpreted this behavior as an example that the fragmentation is not a self-organizing phenomenon, in contrast to the assumption made in a number of papers beginning with the well-known paper [25]. Generally speaking, it casts doubt on the critical nature of fragmentation, which was discussed in Refs. [7, 12, 13, 15-17].

Independence of the exponent $\tau$ on imported energy has been found in the studies [9, 13,16, 17, 25]. At the same time the dependence of $\tau$ on fracture criterion [9] or on constitutive equation of the material [17] was shown. Moreover, it was stated in Refs. [16, 18] that the apparent dependence of τ on the imported energy found in a number of works was related to misinterpretation of the results of measurements or numerical simulations.

A distinctive feature of the present work is that the fragmentation has been numerically simulated using the complete system of equations of deformed solid mechanics (DSM) by the method of smoothed particle hydrodynamics (SPH) [27–29] in a three-dimensional (3D) setting. This approach, which has been previously used in Refs. [8, 9], allows one to verify conclusions based on molecular dynamics and discrete-element modeling widely used for solving the problems of fragmentation [12, 13, 16-18, 24].

When the projectile is fragmented on a thin continuous plate, a high degree of similarity of the simulation results is observed [8]. It is of interest to investigate an analogous problem for mesh shields. However, unlike the plate, an additional length parameter appears here, since the mesh cell is characterized by two parameters-the diameter of the wire and the aperture (the distance between the wires that is visible at the gap). The appearance of an additional parameter substantially complicates the problem; in particular, it requires more calculations.

In the present work we consider the problem of the fragmentation of aluminum projectile on a thin steel mesh shield at high-velocity impact. Quantitative characteristics of the projectile fragmentation are obtained by studying statistics of the cloud of fragments. The considerable attention is given to scaling laws accompanying the fragmentation of the projectile.

## 2. Numerical simulation method and material model

### 2.1. Numerical simulation method.

Numerical simulations in 3D geometry were based on the complete system of equations of DSM and performed using the SPH method implemented in LS-DYNA Version 971 program package [30]. Developed at the Livermore National Laboratory (United States), this software provides solution of 3D dynamic nonlinear problems of deformed solid mechanics. The LS-DYNA Version 971 program package includes SPH algorithm. An exhaustive review of SPH theory and application can be found in [27-29].

The simulations were performed for the spherical aluminum-alloy projectiles with the diameters of 6.35, 7.92 and 9.5 mm and the steel meshes with wire diameter $d_w$ = 0.6 mm and mesh cell apertures $l_a$ from 0.6 mm to 2.2 mm. In all calculations the projectile motion line was perpendicular to the shield plane and was aimed at the node (intersection of wires) located in the center of the mesh shield. The number of SPH particles used by us in calculations for different

projectile–shield pairs are presented in Table 1. The number of SPH particles in the shield depends on the geometric parameters of the mesh. To estimate the total number of particles, we showed in Table 1 the number of SPH particles for the shield that correspond to the mesh with parameters $l_a \times d_w$ = 2.0 mm x 0.6 mm.

**Table 1**. The number of SPH particles used in calculations.

| The projectile diameters $D_{prj}$ (mm) | 6.35 | 7.92 | 9.50 |
|---|---|---|---|
| The number of SPH particles in the projectile | 17269 | 35825 | 59757 |
| Total number of SPH particles in the projectile and shield | 27979 | 54455 | 78387 |

In this paper, results of simulations are only presented for fragments of the projectile. This is related not only to the general pattern of fragmentation, which shows that the characteristic fracture times of projectile and shield are significantly different [8]. The problem of projectile fragmentation at high impact velocity on a shield is also related to the task of spacecraft protection from meteoroids and space debris. As is known [11, 21], this protection is usually based on a scheme, whereby a thin shield is placed in front of a wall to be protected. Fragmentation of a projectile (modeling a meteoroid or space debris) on the shield at high-impact velocities typical of space conditions leads to redistribution of the impact momentum over a larger area, thus reducing the probability of wall perforation. Therefore, the issue of projectile fragmentation is traditionally topical of spacecraft engineering.

In all simulations, the calculations were performed up to the point in time when the distribution of fragments by mass could be treated as stationary. Depending on the size of the projectile, this time was $t_{st}$ = 50 ÷ 70 μs after the impact. The initial data for the program of fragments search were the 3D coordinates of all SPH particles at the time $t_{st}$. Another important characteristic in the search for fragments is the radius of influence $r_{inf}$ that has the meaning of a maximum distance at which two particles can occur so as to belong to the same fragment. It should be recalled that, in SPH calculations, the initial conditions are usually set so that the SPH particles are located at vertices of a cubic lattice [27-30]. Following [9], the radius of influence in the present simulations was selected to be $r_{inf} = a\sqrt{3}$, where $a$ is the cubic lattice constant.

The results of the simulations were averaged over an ensemble (i.e., no less than 10) of simulations corresponding to the same value of the impact velocity. Average values of fragment masses, cumulative distributions, etc. were calculated. The average cumulative distribution was constructed from the averaged differential distribution. Calculations corresponding to the same impact velocity $V$ differed from each other by perturbations introduced into the initial conditions through angular displacement of the projectile relative to its axis of rotation.

**2.2. Material models**

The material models are similar to those used in Ref. [8]. Plastic flow regime is governed by the Prandtl-Reuss flow rule with the von Mises yield condition [30]. Mie-Gruneisen equation of state [31] and Johnson-Cook model [32] for the yield strength were taken as the constitutive equations. Data for the aluminum alloy and steel that we used in the calculations are shown in Table. 2, where they are specified with: $\rho_0$ - initial density of the material, $K$- bulk modulus, $G$ - shear modulus, $\sigma_p$ - tensile strength, $k$- factor in the shock adiabat $D=c_0+k*U$, $\Gamma$- Gruneisen coefficient, which was assumed to be constant, $T_m$ - melting temperature. In Table 2 the values $\sigma_p$ were taken from the spall strength measurements for the aluminum alloy AMG6 and the steel St.3 from Ref. [33]. Parameters for the Johnson-Cook model were taken from Ref. [32].

**Table 2.** Material parameters

| Material | $\rho_0$, g/cm$^3$ | $K_s$, GPa | $G$, GPa | $\sigma_p$, GPa | $k$ | $\Gamma$ | Heat capacity kJ/(kg·°K) | $T_m$, °K |
|---|---|---|---|---|---|---|---|---|
| Steel | 7.85 | 166.7 | 76.9 | 1.66 | 1.49 | 1.93 | 0.477 | 1793 |
| Aluminum alloy | 2.71 | 72.8 | 27.3 | 1.15 | 1.34 | 2.0 | 0.875 | 875 |

The fragmentation of the projectile in the range of impact velocities from 2 to 4 km / sec is considered in the present work. At impact velocities of more than 4.5 km/s [8], the melting of the projectile begins and other, more complicated equations of state to simulate the collision process of the projectile and the shield are required.

## 3. NUMERICAL SIMULATION RESULTS

### 3.1 Patterns of spherical projectile fragmentation on thin mesh shield

Simulations (e.g., see Fig. 3 in [9]) and experiments [2, 5, 11] show that the character of projectile fragmentation upon impact on thin shield depends on the impact velocity. It is well known from ballistics data that a projectile perforates a thin shield at much lower velocities than those necessary for the projectile fragmentation. At somewhat lower impact velocities, the shield is perforated and the projectile is partly fractured, but the mass of minor spalled fragments of the projectile is much smaller than the mass of the main (dominating) fragment. This regime is conventionally referred to as damage of the projectile. At higher impact velocities, the projectile exhibits total disintegration and the mass of largest fragment becomes significantly smaller than the initial projectile mass, which is referred to as the regime of fragmentation. Thus, it can be suggested that a critical impact velocity $V_c$ exists, above which ($V > V_c$) the fragmentation takes place.

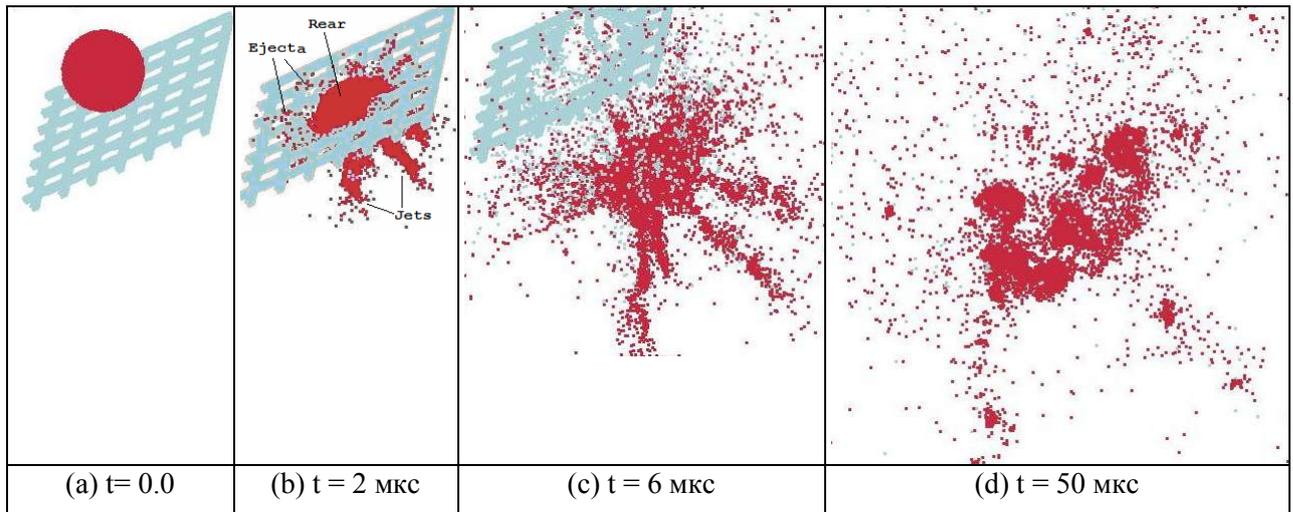

| (a) t= 0.0 | (b) t = 2 мкс | (c) t = 6 мкс | (d) t = 50 мкс |

**Fig.1** (a-d). Typical pattern upon the impact of aluminum projectile (6.35 mm diameter) at 3 km/s velocity on steel mesh ($l_a$ x $d_w$ =2.0 mm x 0.6 mm). All figures are given on the same scale, for this reason Fig. 2 (d) shows only the central part of the fragments cloud.

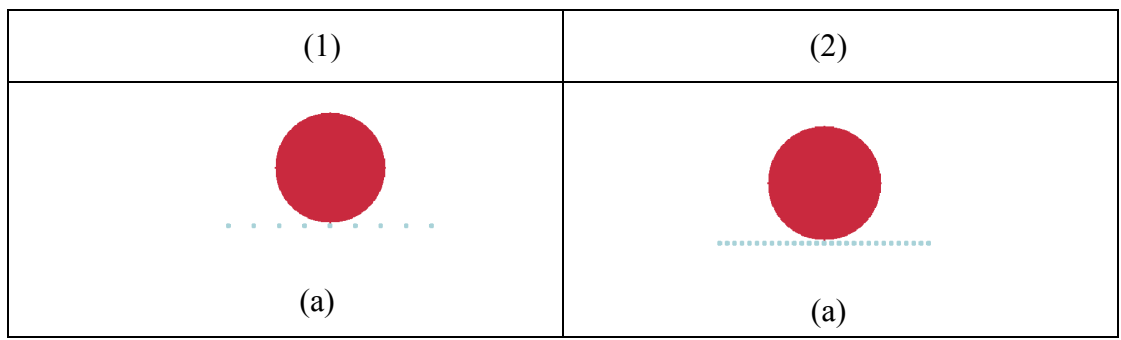

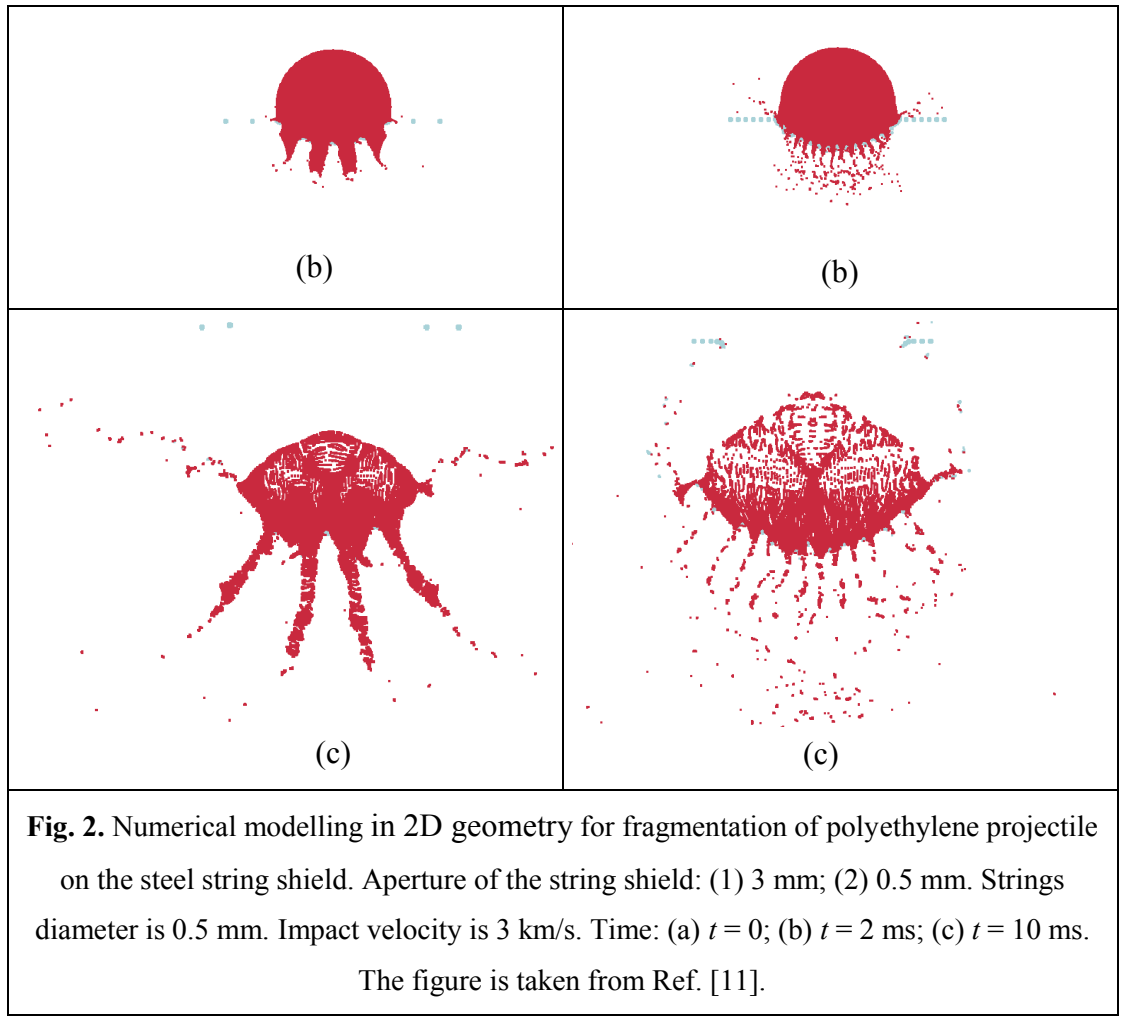

**Fig. 2.** Numerical modelling in 2D geometry for fragmentation of polyethylene projectile on the steel string shield. Aperture of the string shield: (1) 3 mm; (2) 0.5 mm. Strings diameter is 0.5 mm. Impact velocity is 3 km/s. Time: (a) $t = 0$; (b) $t = 2$ ms; (c) $t = 10$ ms. The figure is taken from Ref. [11].

The simulation of the interaction of an aluminum projectile with a steel mesh at a velocity exceeding the critical one shows (Fig. 1) three main properties of the fragments cloud. Firstly, particles are ejected in the direction opposite to the direction of impact. This is so-called ejecta (Fig. 1b), whose mass is much less than the mass of the ejecta in the case of an impact on continuous shield. This observation, made on the basis of calculations, is confirmed experimentally [22]. Secondly, the projectile fragmentation is characterized by forming jets ejected from the front part of the projectile along and across its movement direction (Fig. 1c). Thirdly, the rear part of the projectile at long evolution times of the fragments cloud breaks up into several fragments, which are the largest fragments in the cloud (Fig. 1d). The last two properties are completely different from those observed in the interaction of the projectile with the continuous shield (see Fig. 1 of Ref. [8] or experimental works [5, 6]).

However, as is shown by experiments and simulations [11] with meshes of different apertures, the fragmentation of the projectile on the mesh shields is characterized by both forming frontal jets and shock-wave fracture of the projectile rear part. The latter is inherent to projectile fragmentation on the continuous shield.

After the contact of the projectile with the mesh wires the shock waves propagate from each wire deep into the projectile. After a while the shock waves merge together forming a joint shock wave. This shock wave is similar to that formed at projectile interaction with the continuous shield; however its intensity strongly depends on the mesh aperture: the intensity of the shock wave grows with decreasing aperture (the limiting case of zero apertures corresponds to the continuous shield). The joint shock wave is reflected from the free rear surface of the projectile that generates fracture of its rear part. That is visually demonstrated in Fig. 2 (1c) and (2c). The fragmentation of the projectile rear part in Fig. 2(2c) with aperture of 0.5 mm is similar to the fragmentation observed on the continuous shield.

Therefore, the frontal fragmentation dominates at larger mesh apertures. But when the aperture lessens the part of projectile mass fragmented due to jets forming diminishes significantly and the shock-wave fragmentation of the projectile dominates. Which mode (mechanism) of the fragmentation dominates depends on ratio of the projectile diameter to the mesh period

$$K = D_{prj}/(l_a + d_w), \qquad (2)$$

and, generally speaking, depends on the ratio of the mesh cell aperture to the wire diameter

$$\kappa = l_a / d_w. \qquad (3)$$

The first parameter $K$ defines the number of the mesh cells $(l_a + d_w)$ falling within the projectile diameter, the second parameter $\kappa$ characterizes the shield discreteness (for example, there is no spacing between the wires when $\kappa = 0$, and the discrete shield may be regarded as a continuous one). Thus, in the case of the projectile fragmentation on the mesh shield, the solution depends on two dimensionless geometric parameters (2) and (3). In the present work, the results of simulations are presented for a fixed wire diameter of 0.6 mm. Therefore, we consider the solution depending on one dimensionless geometric parameter (2).

Experiments on the interaction of the projectile with a continuous plate [5,6] with analogous $h/D_{prj}$ ratios ($h$ is the thickness of the plate) and close impact velocities showed that the morphology and internal structure of the clouds of fragments remain similar. This property is used in testing the shield protections intended for installation on the spacecraft (see, for example, [34]). For mesh shields, such a generally recognized similarity parameter, analogous to $h/D_{prj}$ for continuous shields, has not yet been revealed. In fact, the work presented, following our experiments [11], tests the parameter (2) for this role.

**3.2 Average fragment mass as function of impact velocity for various mesh shields.**

As was noted above, the fragmentation of a projectile requires that the impact velocity $V$ would exceed the critical value ($V_c$). For quantitative characterization of the degree of fracture, it is a common practice to calculate the average fragment mass as $M_{avr} = \langle \overline{M}_2^j / \overline{M}_1^j \rangle$, where $\overline{M}_1^j$ and $\overline{M}_2^j$ are the first and second moments of fragment mass distribution (4) in the $j$th simulation, where angle brackets $\langle \ldots \rangle$ denote averaging over the ensemble (i.e., no less than 10) simulations for the same value of impact velocity $V$. As was noted above, simulations with the same value of impact velocity $V$ differed from each other by the angular perturbation introduced into initial conditions through angular displacement of the projectile relative to its axis of rotation. The $k$th moments $\overline{M}_k^j$ of fragment mass distribution in the $j$th simulation is defined in a single fragmentation event as [3, 12, 19, 20]

$$\overline{M}_k^j(V) = \sum_m {}^{\wedge} m^k n^j(m, V), \qquad (4)$$

where $n^j(m, V)$ is the number of fragments with mass $m$ in the $j$th simulation at velocity $V$. Here, the cap over sum in Eq. (3) denotes that the sum runs over all fragments, excluding the largest one produced in the event.

Fig. 3 shows the dependence $M_{avr} / m_{tot}$ ($m_{tot}$ is the projectile mass) on the impact velocity $V$ for the projectile diameter $D_{prj} = 6.35$ mm and different mesh shields with the aperture $l_a$ of 0.0 to 2.0 mm and the wire diameter $d_w = 0.6$ mm (aperture $l_a = 0.0$ corresponds to the continuous shield). The peaks of the dependences correspond to the critical impact velocities $V_c$, at which fragmentation occurs [8].

Fig. 4 shows the envelope of the peak values $(M_{avr})_{max} / m_{tot}$ of the average fragment mass presented in Fig. 3 for projectile diameter $D_{prj} = 6.35$ mm, and for two other projectile diameters depending on the parameter $K$, which defines the number of the mesh periods falling within the projectile diameter (2). It is seen that everywhere $(M_{avr})_{max} / m_{tot}$ decreases non-monotonically with increasing parameter $K$ (by decreasing the aperture of the mesh cell), passing through the local minimum. The presence of a local minimum appears to be due to the two fragmentation modes that occur with "small" and "large" aperture values (see Section 3.1).

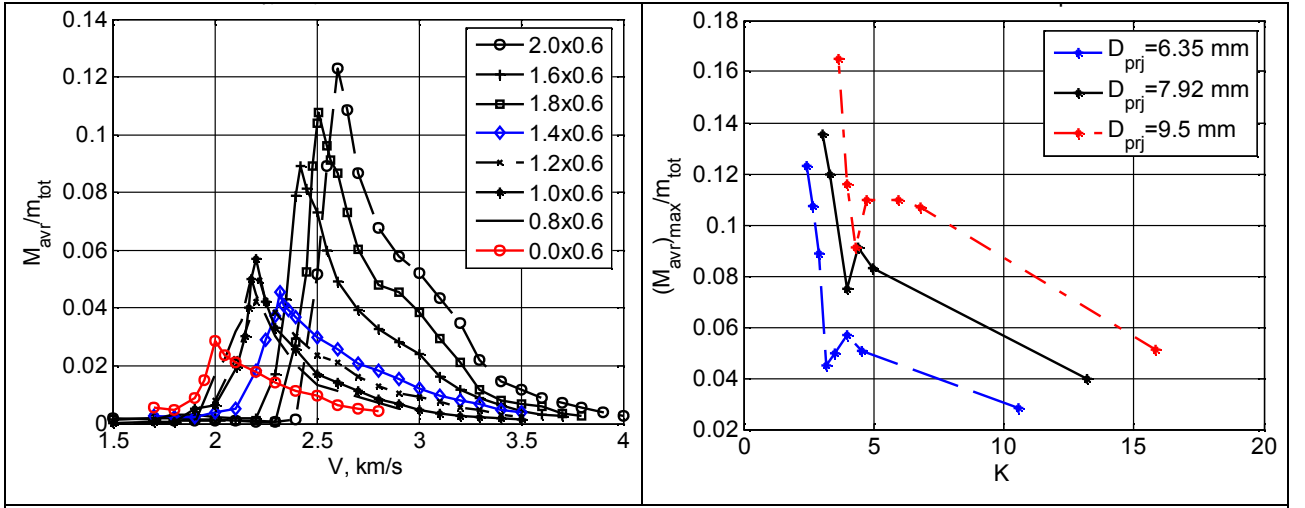

**Fig. 3.** Normalized average fragment mass ($M_{avr}/m_{tot}$) vs. impact velocity $V$ for the 6.35 mm diameter projectile and the mesh shields having different parameters $l_a \times d_w$ (shown in the figure in mm). In figure $m_{tot}$ is the projectile mass.

**Fig. 4.** Envelope of the peak values of the average fragment mass (($M_{avr})_{max}/m_{tot}$) vs. parameter $K$ (2) for different diameters of the projectile (shown in the figure).

In order to obtain a criterion separating "small" and "large" apertures, we proceed as follows. Let us change the scale of the curves in Fig. 4 along the abscissa axis by using the substitution $D_{prj}/(l_a+d_w) \to (D_{prj})^\gamma/(l_a+d_w)$, where $\gamma$ is the fitting parameter. The best coincidence of the local minima of the curves is obtained for the value $\gamma = 1/4$, thus we obtain $(D_{prj})^{1/4}/(l_a+d_w) = 0.8$ (mm$^{-3/4}$). Hence the critical value of the mesh aperture is

$$(l_a)_c = 1.25(D_{prj})^{1/4} - d_w, \tag{5}$$

where the projectile diameter $D_{prj}$ and the wire diameter $d_w$ must be taken in mm. Thus, we will assume that the values $0 < l_a < (l_a)_c$ correspond to "small" apertures, and $l_a > (l_a)_c$ - "large" apertures. For the projectile diameters of 6.35, 7.92 and 9.5 mm, we have from (5) $(l_a)_c \approx$ 1.4, 1.5 and 1.6 mm, respectively.

### 3.3 Critical impact velocity as function of mesh aperture and projectile diameter.

In the general case, the dependences of the normalized average fragment mass $M_{avr}/m_{tot}$ on the impact velocity $V$ shown in Fig. 3, do not have scale invariance. However, for two groups of curves corresponding to "small" and "large" apertures, scaling still takes place.

Fig. 5 shows that a collapse of the three curves taken from Fig. 3 (corresponding to the "large" apertures $l_a$ = 1.6, 1.8 and 2.0 mm) can be obtained. The data presented in Fig. 5 is scaled with respect to coordinate axes by means of the transformation $V \to VK^\alpha$ and $M_{avr}/m_{tot} \to (M_{avr}/m_{tot})K^\beta$. The best result is obtained for the exponents α = 0.4 and β = 1.70. The collapse implies that $(M_{avr}/m_{tot})K^\beta$ is only the function of $VK^\alpha$. From Fig. 5 we obtain the relation $V_c K^{0.4}$ = 3.70 km/sec for the critical impact velocity. From this, taking into account (2), we have the dependence of the critical velocity $V_c$ on the aperture in the region of "large" apertures:

$$V_c = 3.7 K^{-0.4} \quad \text{or} \quad V_c = k_{1a}(l_a + d_w)^{0.4}, \tag{6}$$

where $k_{1a}$ = 1.8 km/sec·(mm)$^{-0.4}$. It can be seen that the critical velocity increases with increasing mesh aperture.

For the curves in Fig. 3, corresponding to the apertures $l_a$ = 1.2, 1.0, 0.8 and 0.0 mm, scale transformations allow one to match the maxima of the curves only on the abscissa axis as shown in Fig. 6. The best coincidence is obtained for the exponent α = 0.095. As a result we have the relation $V_c K^{0.095}$ = 2.5 km/sec and the dependence of the critical velocity on the parameter $K$ for "small" apertures in the form

$$V_c = 2.5 K^{-0.095} \quad \text{или} \quad V_c = k_{2a}(l_a + d_w)^{0.095}, \tag{7}$$

where $k_{2a}$ = 2.1 km/sec·(mm)$^{-0.095}$. It is seen that in the case of "small" apertures, the critical velocity grows very slowly with increasing mesh aperture.

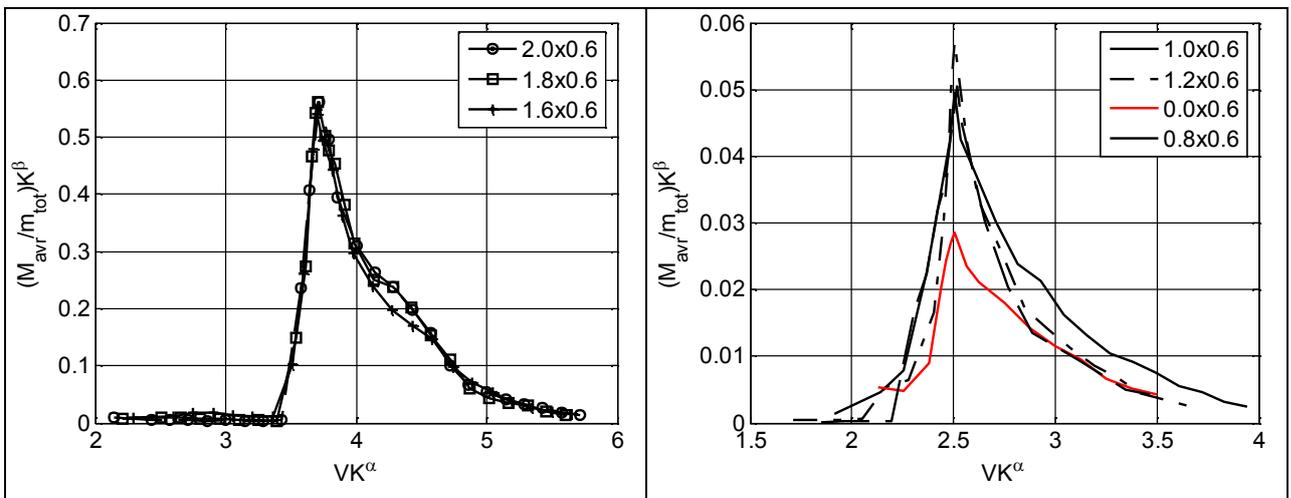

**Fig. 5.** Collapse of the three curves taken from Fig. 3 (the "large" apertures $l_a$ = 1.6, 1.8 and 2.0 mm) obtained by rescaling the two axes with scaling exponents α = 0.4 and β = 1.70.

**Fig. 6.** Rescaling the abscissa axis with appropriate power of parameter *K* for the four curves taken from Fig. 3 (the "small" apertures $l_a$ = 1.2, 1.0, 0.8 and 0.0 mm) with scaling exponent α=0.095 (β=0).

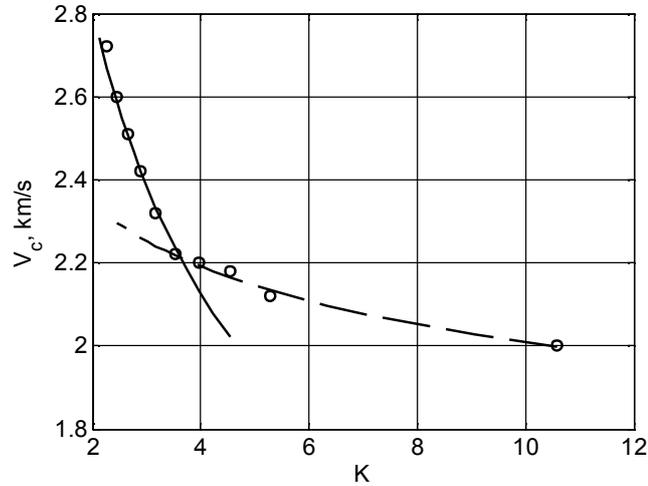

Fig. 7. Comparison of the critical impact velocities obtained by the numerical simulation (o) and with the help of formulas (6) (—) and (7) (- - -) for the 6.35 mm diameter projectile.

Fig. 7 shows the dependence of the critical velocity $V_c$ on the dimensionless parameter *K* (2) obtained by numerical simulation with the 6.35 mm diameter projectile. In the figure, the curves of functions (6) (—) and (7) (- - -) are also drawn for comparison. On the upper branch (—) corresponding to the "large" apertures, besides the points corresponding to the mesh apertures $l_a$ = 1.6, 1.8 and 2.0 mm (by which the dependence (6) was constructed), the points corresponding to the apertures $l_a$ = 1.2, 1.4 and 2.2 mm are also plotted. On the lower branch (- - -) corresponding to small apertures, the points with apertures $l_a$ = 1.2, 1.0, 0.8, 0.6 and 0.0 mm are located. The extreme right point on the lower branch ($l_a$ = 0.0 mm) corresponds to a continuous plate.

The presence of a sharp bend (Fig. 7) in the dependence of the critical velocity on the parameter *K* (2) is associated with two fragmentation modes (see Section 3.1), which take place for "small" and "large" values of the mesh apertures.

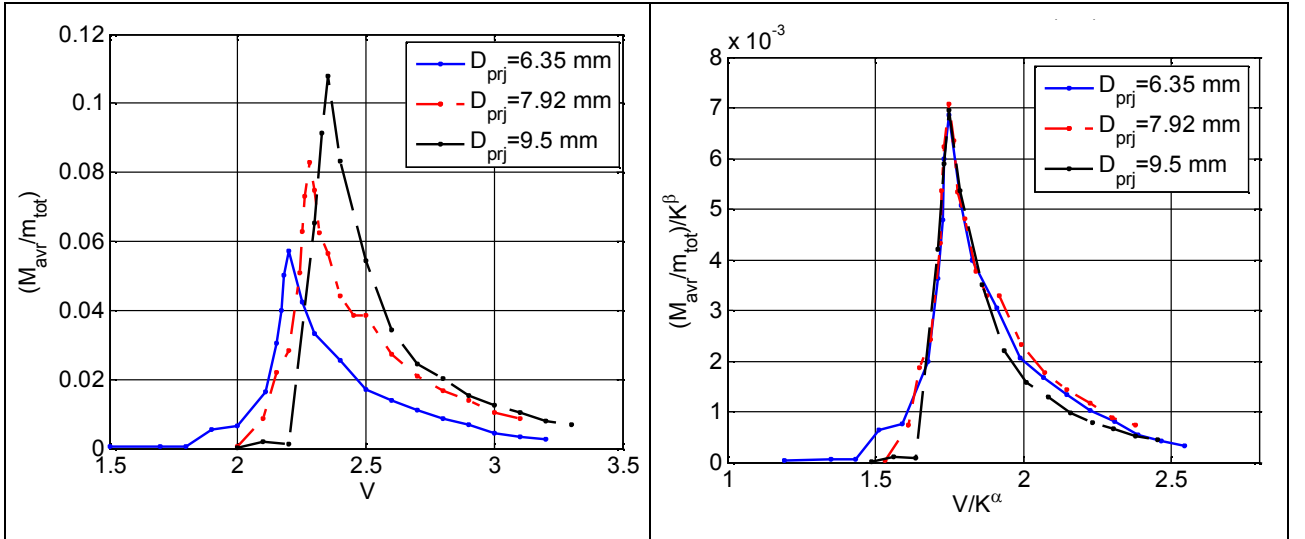

**Fig. 8. (a)** Dependences of the normalized average mass $M_{avr}/m_{tot}$ of the fragments on the impact velocity $V$ for the mesh shield with $l_a \times d_w$ = 1.0 mm x 0.6 mm and the projectiles of different diameters (shown in the figure).
  **(b)** Rescaling the two axes with appropriate powers of the parameter $K$ for the data taken from Fig. 8(a) with scaling exponents $\alpha$ = 0.166 and $\beta$ = 1.54.

Let us consider the dependence of the critical velocity on the projectile diameter with the invariable mesh shield for the cases of "small" and "large" apertures.

Fig. 8 (a) shows the dependence of the normalized average mass $M_{avr}/m_{tot}$ of the fragments on the impact velocity $V$ for the mesh shield with parameters $l_a \times d_w$ = 1.0 mm x 0.6 mm and the projectiles with diameters of 6.35, 7.92 and 9.5 mm. The next Fig. 8 (b) shows the collapse of these dependencies, which is obtained by rescaling the two axes by means of the transformations $V \rightarrow V/K^\alpha$ and $M_{avr}/m_{tot} \rightarrow (M_{avr}/m_{tot})/K^\beta$. The best result is obtained for the exponents $\alpha$ = 0.166 and $\beta$ = 1.54.

From Fig. 8(b) we obtain the relation $V_c/K^{1/6}$ =1.75 km/sec for the critical impact velocity. From this, taking into account (2), we have the dependence of the critical velocity $V_c$ on the projectile diameter in the region of "small" apertures:

$$V_c = k_{2d}(D_{prj})^{1/6} \qquad (8)$$

where $k_{2d}$ = 1.6 km/sec·(mm)$^{-1/6}$. It is seen that in the case of "small" apertures, the critical velocity grows with increasing the projectile diameter.

Combining (7) and (8), we obtain the dependence of the critical velocity on the projectile diameter and the mesh parameters for "small" apertures in the form

$$V_c \sim (D_{prj})^{1/6} \cdot (l_a + d_w)^{0.095} \qquad (9)$$

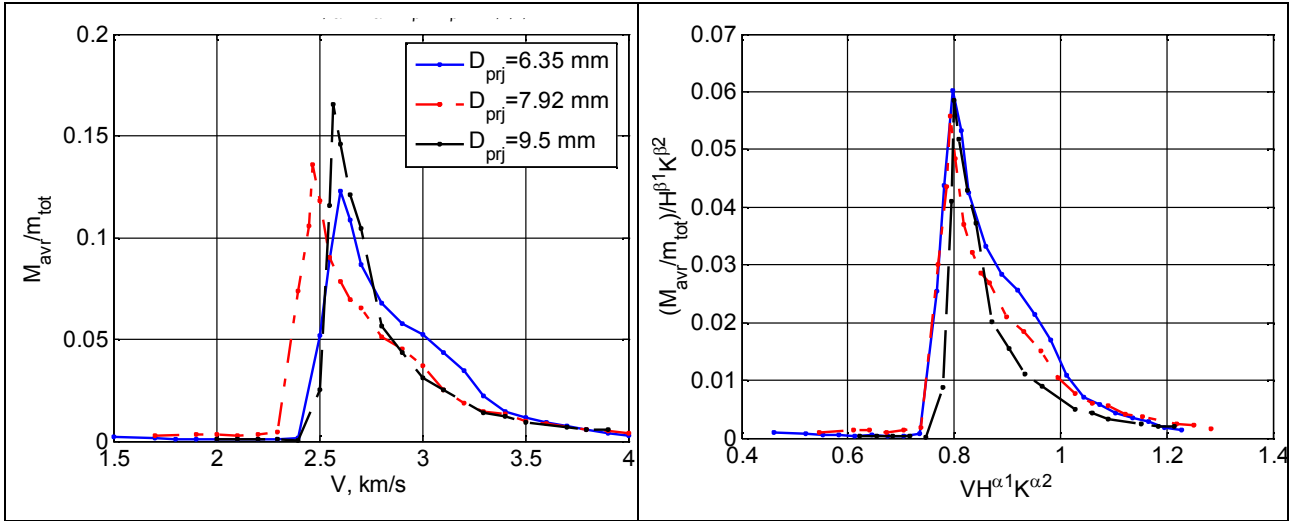

**Fig. 9 (a)** Dependences of the normalized average mass $M_{avr}/m_{tot}$ of the fragments on the impact velocity $V$ for the mesh shield with $l_a \times d_w$ = 2.0 mm x 0.6 mm and the projectiles of different diameters (shown in the figure).

**(b)** Rescaling the two axes with appropriate powers of the parameters $K$ and $H$ for the data taken from Fig. 9(a) with scaling exponents $\alpha_1 = 0.75$, $\alpha_2 = 0.70$, $\beta_1 = 0.0$ and $\beta_2 = 0.8$.

Fig. 9 (a) shows the dependence of the normalized average mass $M_{avr}/m_{tot}$ of the fragments on the impact velocity $V$ for the mesh shield with parameters $l_a \times d_w$ = 2.0 mm x 0.6 mm and the projectiles with diameters of 6.35, 7.92 and 9.5 mm. The scale transformations allow one to coincide only the peaks of the curves as shown in Fig. 9 (b). The scale transformations found have the form $V \rightarrow VH^{\alpha_1}K^{\alpha_2}$ and $M_{avr}/m_{tot} \rightarrow (M_{avr}/m_{tot})/H^{\beta_1}K^{\beta_2}$, where the parameter $K$ is taken from (2), the parameter $H = m_{eff}/M_{prj}$, where $m_{eff}$ is the effective mass of the shield shadowed by the incident projectile [25]. For mesh shield $l_a \times d_w$ = 2.0 mm x 0.6 mm the calculations yield $H = 0.090$, 0.078 and 0.063 for the projectile diameters of 6.35, 7.92 and 9.5 mm, respectively. The best coincidence shown in Fig. 9(b) is obtained for the exponents $\alpha_1 = 0.75$, $\alpha_2 = 0.70$, $\beta_1 = 0.0$ and $\beta_2 = 0.8$. Therefore, from Fig. 9(b) we obtain the relation $V_c H^{0.75} K^{0.70} = 0.8$ km/sec and the dependence of the critical velocity on the parameters $K$ and $H$ for "large" apertures in the form:

$$V_c = k_{1d} H^{-0.75} (D_{prj})^{-0.70} \qquad (10)$$

where $k_{1d}$ = 1.56 km/sec·(mm)$^{0.7}$. It is seen that in the case of "large" apertures, the critical velocity increases with increasing diameter of the projectile as $\sim (D_{prj})^{1.55}$ and decreases with increasing effective mass of the mesh shadowed by the incident projectile as $\sim (m_{eff})^{-0.75}$.

From (7) and (10) we obtain the dependence of the critical velocity on the projectile diameter and the mesh parameters for "large" apertures in the form

$$V_c \sim (m_{eff})^{-0.75} (D_{prj})^{1.55} (l_a + d_w)^{0.4} \qquad (11)$$

### 3.4. Weight-average fragment mass as function of impact velocity.

Let us again turn to calculation of the average fragment mass, but this time with allowance for the largest fragment, and define the weight-average fragment mass as follows:

$$m_{avr} = \frac{\langle M_2^j(V) \rangle}{M_1(V)}, \qquad (12)$$

Here, the first ($M1$) and second ($M2$) moments are defied as

$$M_k^j(V) = \sum_m m^k n^j(m, V). \qquad (13)$$

where notation in the right-hand part is the same as in Eq. (4). The difference between Eqs. (4) and (13) is that in (13) the sum is taken over all fragments including the largest one. In this case, the first moment represents the total mass of fragments, which remains unchanged; $M_1$ equals the initial projectile mass $m_{tot}$ and can be removed from averaging operator in Eq. (12).

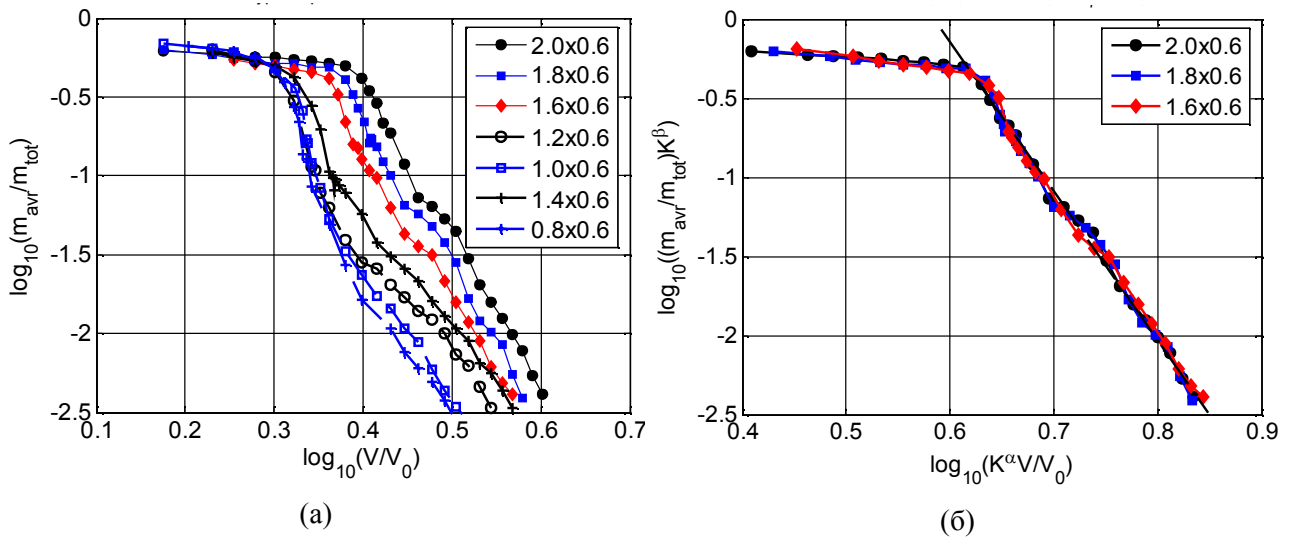

(a)  (б)

**Fig. 10**. The normalized weight-average fragment mass ($m_{avr} / m_{tot}$) vs. impact velocity $V$ for the projectile diameter $D_{proj} = 6.35$ mm and the mesh shield with different parameter $l_a \times d_w$ (shown in the figure in mm). $V_0 = 1$ km/sec.

**Fig. 11**. Rescaling the two axes with appropriate powers of the parameter $K$ for the three meshes (shown in figure in mm) taken from Fig. 10 with scaling exponents $\alpha = 0.6$ and $\beta = 0.0$.

Fig. 10 shows the normalized weight-average fragment masses, depending on the impact velocity for the projectile diameter $D_{proj}$ = 6.35 mm and the various mesh shields. For the dependences in Fig. 10 only simulation results for meshes with the "large" aperture of 2.0 mm, 1.8 mm, 1.6 mm can be match by means of scale transformations. The result of rescaling is shown in Fig. 11. It can be seen that the rescaling the two axes with appropriate powers of the parameter $K$ gives collapse of the curves with scaling exponents $\alpha$ = 0.6 and $\beta$ =0.0. Thus, here the scale invariance occurs only for the normalized weight-average fragment masses of the meshes with "large" apertures.

In the supercritical region ($V > V_c$), the data in Fig. 11 is approximated (with the of least square method) by a straight line. As a result, we have obtained that for $V > V_c$ in the region of "large" apertures, the weighted average mass is

$$m_{avr} \propto \left( \left( \frac{l_a + d_w}{(D_{prj})_0} \right)^{-0.6} \frac{V}{V_0} \right)^\kappa, \quad (14)$$

were $\kappa$ = -9.4±0.2, $V_0$ = 1 km/s and $(D_{prj})_0$ = 6.35 mm.

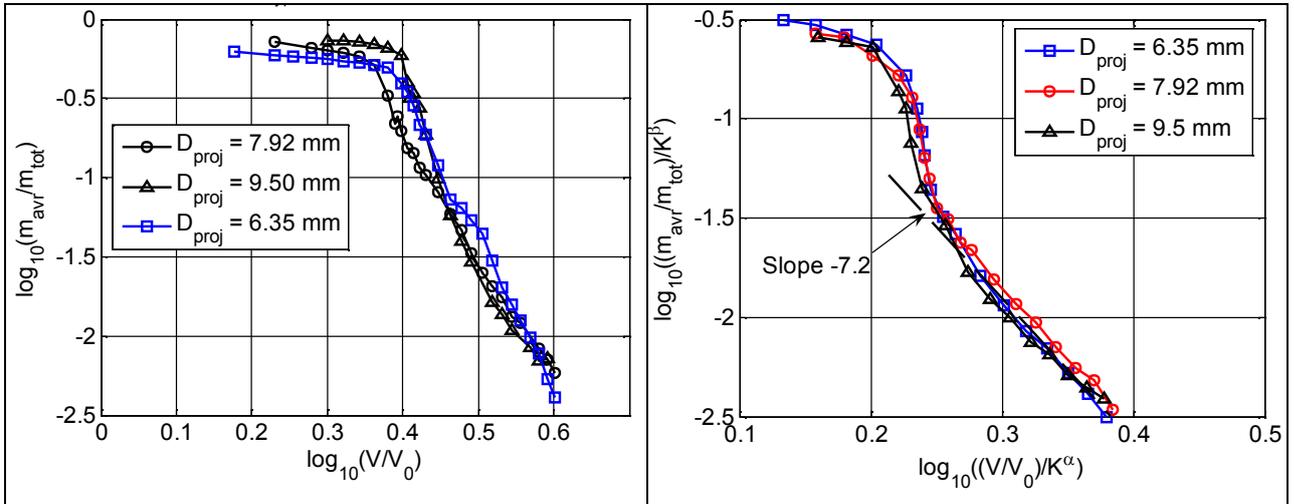

**Fig. 12**. The normalized average fragment mass ($M_{avr}/m_{tot}$) vs. impact velocity $V$ for the mesh shield with $l_a \times d_w$ = 2.0 mm x 0.6 mm and the projectiles of different diameters (shown in the figure). $V_0$=1 км/сек.

**Fig. 13**. The normalized average fragment mass ($M_{avr}/m_{tot}$) vs. impact velocity $V$ for the mesh shield with $l_a \times d_w$ = 1.0 mm x 0.6 mm after rescaling of the two axes with appropriate powers of the parameter $K$. The figure shows the best coincidence of the curves, which is observed with scaling exponents $\alpha$ =0.25 and $\beta$ = 0.8.

Fig. 12 shows the normalized weight-average fragment masses, depending on the impact velocity for the projectile of various diameters and mesh with $l_a \times d_w$ = 2.0 mm x 0.6 mm. It can be seen that in the supercritical region ($V > V_c$) the data corresponding to different sizes of the projectiles do not have a universal asymptotic behavior in the form $m_{avr} \propto V^\chi$. The power exponent $\chi$ depends on the projectile diameter, varying from -8.2 to -9.4.

In contrast, in the region of "small" apertures for mesh with $l_a \times d_w$ = 1.0 mm x 0.6 mm, it is possible to obtain good coincidence of the curves by rescaling the two axis with help of the parameter $K$ (2). Fig. 13 shows the collapse of the curves, which is observed for the exponents $\alpha$ = 0.25 and $\beta$ = 0.8. It can be seen that in the supercritical region ($V > V_c$), the data corresponding to different diameters of the projectile have approximately the same asymptotic form

$$m_{avr} \propto D^{0.8-0.2\chi}(V/V_0)^\chi \tag{15}$$

with $\chi$ = -7.2.

### 3.5. Average cumulative fragment mass distributions.

The mass distribution of the fragments is described by cumulative distribution

$$N(m) = \int_m^\infty n(m')dm', \tag{16}$$

where $n(m)$ is the number of the fragments with mass $m$. For a given impact velocity $V$, the average cumulative distribution $<N(m,V)>$ was constructed using the differential distribution obtained by averaging over a series of no less than ten simulations with the same $V$ value.

As shown in experiments and numerical simulations, [1, 3, 7, 10, 12-16, 19, 23-25] $n(m)$ has the form of the power function (1) at the critical and above the critical point $V_c$ in some non-negligible range of fragments mass variation. In this range the cumulative distribution is $N(m) \sim m^{1-\tau}$. In many studies, the universality of the exponent $\tau$ was noted. For example, the independence of $\tau$ on the imported energy or the impact velocity was found in Refs. [9, 13, 16, 17, 25]. At the same time, experiments and numerical simulation have shown the dependence of $\tau$ on the effective dimension of the fragmented object [23]. Also, the dependence of the exponent $\tau$ on the failure criterion [9] or on the constitutive equation of the material [17] was shown.

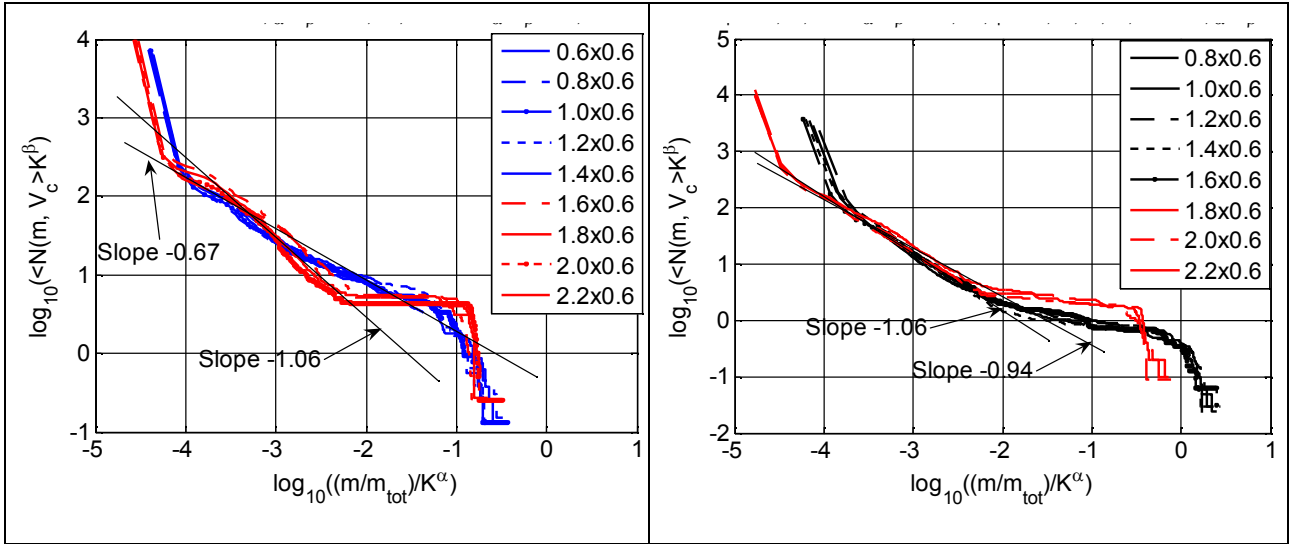

**Fig. 14.** Average cumulative fragment mass distributions at critical impact velocities for the projectile diameter $D_{prj}$ = 6.35 mm and the meshes with parameters $l_a \times d_w$ shown in the figure in mm after rescaling the two axes with help of the parameter $K$ (2). The figure shows the best coincidence of the curves, which is observed with scaling exponents $\alpha$ =0.25 and $\beta$ =0.16 ("small" apertures) and $\alpha$ = 0.8, $\beta$ = 0.9 ("large" apertures).

**Fig. 15.** The same as in Fig. 14, but for the projectile diameter $D_{prj}$ = 9.5 mm. The figure shows the best coincidence of the curves, which is observed with scaling exponents $\alpha$ = - 0.8 and $\beta$ = - 0.85 ("small" apertures) and $\alpha$ = 0.0, $\beta$ = 0.0 ("large" apertures).

Fig. 14 shows the average cumulative fragment mass distributions at the critical impact velocities for the projectile diameter $D_{prj}$ = 6.35 mm and the meshes with apertures $l_a$ from 0.6 to 2.0 mm. It can be seen that after rescaling the two axes with help of the parameter $K$ (2) the data in Fig. 14 are assembled into two groups corresponding to the "small" ($l_a$ = 0.6, 0.8, 1.0, 1.2, 1.4 mm) and "large" ($l_a$ = 1.6, 1.8, 2.0, 2.2 mm) apertures. The figure shows the best coincidence of the curves, which is observed with scaling exponents $\alpha$ =0.25 and $\beta$ =0.16 (group of "small" apertures) and $\alpha$ = 0.8, $\beta$ = 0.9 (group of "large" apertures). Thus, the average cumulative fragment mass distributions at the critical points must obey the scaling rule of

$$<N_c(m,(l_a+d_w)/(D_{prj})_0)>_i = [((l_a+d_w)/(D_{prj})_0]^{\beta_i} \Phi_i\left((m/m_{tot})[(l_a+d_w)/(D_{prj})_0]^{\alpha_i}\right), \quad (17)$$

where $(D_{prj})_0$ = 6.35 mm, $<N_c(m)> = <N(m,V_c)>$ and $\Phi_i$ ($i$ = 1, 2) denotes the scaling function for two group of the distributions. In Eq. (17) we took into account that at the scale transformation in Fig. 14 only the mesh period incoming in the parameter $K$ (2) changes.

It can also be seen (Fig. 14 and 15) that each group of distributions in the intermediate mass region has a power-law distribution with an exponent τ different from that in the other group: τ = 2.06±0.02 (group of "large apertures") and τ = 1.67±0.02 (group of "small apertures").

Recall that the separation of mesh shields into groups of "small" and "large" apertures is associated with various modes of the fragmentation (see Section 3.1), and the critical value of the mesh aperture separating the modes of fragmentation from each other is given by formula (5): for the 6.35 mm diameter projectile $(l_a)_c \approx 1.4$. Therefore, we can state the dependence of the exponent τ on the fragmentation mode.

Fig. 15 shows the same as in Fig. 14, but for the projectile diameter $D_{prj}$ = 9.5 mm. It is also seen that the distributions are assembled into two groups corresponding to "small" ($l_a$ = 0.8, 1.0, 1.2, 1.4, 1.6 mm) and "large" ($l_a$ = 1.8, 2.0, 2.2 mm) apertures (for a 9.5 mm diameter projectile $(l_a)_c \approx 1.6$). The difference in the exponents of power-law distributions is less appreciable here: τ = 2.06±0.02 (group of "small apertures") and τ = 1.94±0.02 (group of "large apertures"). For this case, the scaling in the form (17) also takes place. We note that in Fig. 15 the distributions in the group of "large apertures" initially show good coincidence and do not require scale transformations.

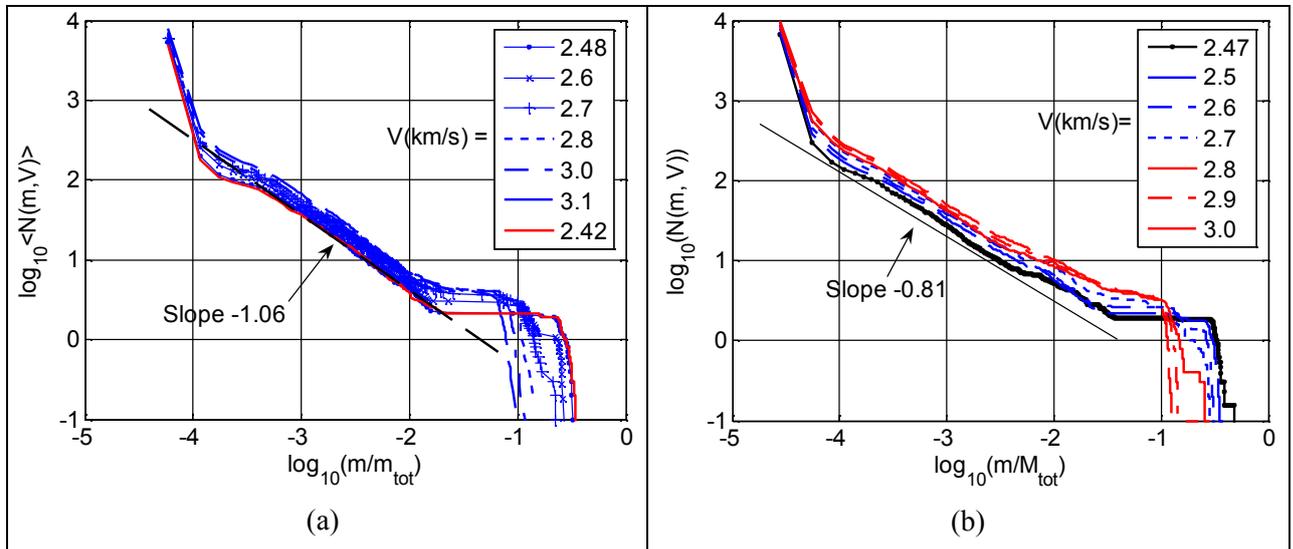

**Fig. 16**. Average cumulative fragment mass distributions at the impact velocities $V \geq V_c$:

a) $D_{prj}$ = 6.35 mm, mesh $l_a \times d_w$ = 1.6 mm x 0.6 mm, $V_c$ = 2.42 km/s;

b) $D_{prj}$ = 7.92 mm, mesh $l_a \times d_w$ = 2.0 mm x 0.6 mm, $V_c$ = 2.47 km/s.

Fig. 16 presents the simulation results of the average cumulative fragment mass distributions at the impact velocities exceeding the critical velocity, $V \geq V_c$, for the projectile diameters $D_{prj}$ =

6.35 and 7.92 mm, and the meshes $l_a \times d_w$ = 1.6 mm x 0.6 mm and 2.0 mm x 0.6 mm, respectively. It can be seen that the exponent of the power-law distribution in the intermediate mass range does not depend on the impact velocity in a wide range of the velocity variations and remains approximately equal to the value of τ at the critical impact velocity. For the projectile diameter $D_{prj}$ = 6.35 mm, τ = 2.06±0.02, and for the projectile diameter $D_{prj}$ = 7.92 mm, τ = 1.81±0.02.

## 4. Summary and conclusions

In the present work we have considered the problem of the fragmentation of aluminum projectile on a thin steel mesh shield at high-velocity impact in a three-dimensional (3D) setting. Quantitative characteristics of the projectile fragmentation were obtained by studying statistics of the cloud of fragments. The considerable attention was given to scaling laws accompanying the fragmentation of the projectile.

In well-known numerical simulations of fragmentation in mechanical systems, the object under consideration was described as a set of either identical particles coupled by a pairwise potential (by analogy with molecular dynamics) or pieces of various shapes linked by weightless coupling elements of various types. The dynamics of fragments (elements), including both translational and rotational motions, was described by a system of Newton's equations. Simulations were most frequently performed by the method of molecular dynamics or discrete element models [12, 13, 16-18].

A distinctive feature of the present work is that here, following our works [8, 9], the fragmentation has been numerically simulated using the complete system of equations of deformed solid mechanics by the method of smoothed particle hydrodynamics in a 3D setting. The behavior of materials was described using the Mie–Gruneisen equation [31] of state and the Johnson–Cook plasticity model [32].

Simulations were performed for a spherical aluminum-alloy projectiles with the diameters of 6.35, 7.92 and 9.5 mm and a steel meshes with wire diameter $d_w$ = 0.6 and cell apertures $l_a$ from 0.6 mm to 2.2 mm. In all simulations the projectile motion line was perpendicular to the shield plane and was aimed at the node (intersection of wires) located in the center of the mesh shield. The number of SPH particles used by us in calculations for different projectile–shield pairs are presented in Table 1.

The main feature of the projectile fragmentation on the mesh shield is the formation of jets of fragments ejected from the front part of the projectile along and across its movement direction. Experiments and simulations [11] showed that the fragmentation of the projectile on the mesh shields is characterized by both forming frontal jets and shock-wave fragmentation of the projectile rear part. The latter is inherent to fragmentation on a continuous shield. Which mode of the fragmentation dominates depends on the geometrical parameters $K = D_{prj}/(l_a + d_w)$ and $\kappa = l_a/d_w$, where $D_{prj}$ is the projectile diameter, $l_a$ is the mesh aperture, $d_w$ is the wire diameter. The frontal fragmentation dominates at larger cell apertures, but when the aperture lessens the part of projectile mass fragmented due to the jets diminishes significantly and the shock-wave fragmentation of the projectile prevails.

In the present work, the results of simulations are presented for a fixed wire diameter $d_w = 0.6$ mm. Therefore, we consider the solution depending on one dimensionless geometric parameter $K$ defining the number of the mesh periods falling within the projectile diameter.

Experiments on the interaction of the projectile with a continuous plate [5,6] with analogous $h/D_{prj}$ ratios ($h$ is the thickness of the plate) and close impact velocities showed that the morphology and internal structure of the clouds of fragments remain similar. This property is used in testing the shield protections intended for installation on the spacecraft (see, for example, [34]). For mesh shields, such a generally recognized similarity parameter, analogous to $h/D_{prj}$ for continuous shields, has not yet been revealed. In fact, the work presented, following our experiments [11], tests the parameter (2) for this role.

The main conclusions of this work can be formulated as follows.

- The results of modeling show that the process of projectile facture depending on the impact velocity can be separated into two stages: damage and fragmentation, with a sharp transition from one to another at a critical impact velocity $V_c$. This conclusion is consistent with the results of modeling performed by methods of molecular dynamics and discrete element models [10, 12, 13, 16-18], and by means of the numerical simulation by SPH method [8, 9].

- The largest value of the average fragment mass $(M_{avr})_{max}$ observed at the critical impact velocity $V_c$ depends nonmonotonically on the parameter $K$, having a local minimum at some value of $K_c$ or aperture $(l_a)_c$ (Fig. 4). It is reasonable to assume that this is due to the presence of two modes of the projectile fragmentation on the mesh shield [11]. For $K < K_c$ ($l_a > (l_a)_c$ - the "large" mesh apertures), the frontal mechanism of the projectile fragmentation inherent to the mesh shield dominates. For $K > K_c$ ($l_a < (l_a)_c$ - "small" mesh apertures), shock-wave fragmentation of the back part of the projectile inherent to on the continuous shield dominates.

- It is shown that the dependences of the normalized average mass $M_{avr}/m_{tot}$ of the fragments on the impact velocity $V$ exhibit the property of scale invariance, when the parameter $K$ varies, in the entire considered range of the mesh apertures. As a result, we obtained the dependences of the critical velocity $V_c$ on the projectile diameter and the mesh parameters for both "small" (9) and "large" apertures (11). The dependence of the critical velocity $V_c$ of fragmentation on the parameter *K* consists of two branches (Fig. 7), experiencing a sharp bend at the point corresponding to the critical aperture of the mesh, again confirming the existence of two modes of the projectile fragmentation on the mesh shield, which are mentioned above.

- Dependences of the normalized weighted average mass $m_{avr}/m_{tot}$ of fragments on the impact velocity exhibit the property of scale invariance when the parameter *K* varies in the region of "large" apertures only for a fixed diameter of the projectile and in the region of "small" apertures only for a fixed mesh aperture. As a result, in the first case, the dependence of $m_{avr}$ on the impact velocity and mesh parameters (14) was extracted, in the second case - the dependence of $m_{avr}$ on the impact velocity and the projectile diameter (15).

- It is shown that the average cumulative mass distributions, constructed at critical impact velocities, $<N(m,V_c)>$, exhibit the property of scale invariance, splitting into two groups of distributions exactly corresponding to the "small" and "large" apertures of the mesh shield (Fig. 14 and 15). In each group, the average cumulative distributions show good coincidence in the entire mass region. It can also be seen (Fig. 14 and 15) that each group of distributions in the intermediate mass region has a power-law distribution with an exponent τ different from that in the other group. As noted above, the separation of the meshes into groups of "small" and "large" apertures is associated with various modes of fragmentation that are observed when the projectile is disintegrated. Therefore, we can state the dependence of the exponent τ on the fragmentation mode.

**Acknowledgments**

This research was supported by the Russian Foundation for Basic Research (grant 15-01-00565-a).